\documentclass[aps,prb,twocolumn,superscriptaddress]{revtex4-2}
\usepackage{graphicx}
\usepackage{epstopdf}
\usepackage{amsmath}
\usepackage{appendix}
\usepackage{color}
\usepackage[colorlinks,linkcolor=blue,anchorcolor=blue,citecolor=blue,urlcolor=blue]{hyperref}
\usepackage{orcidlink}
\usepackage{mathtools} 
\usepackage{amssymb}
\usepackage{amsthm}

\begin{document}

\title{Prediction of 1:1 Kagome Metals with Superconductivity and Nontrivial Band Topology}

\author{Na Jiao} \email[E-mail: ]{j$\_$n2013@126.com} 
\affiliation{School of Physics and Physical Engineering, Qufu Normal University, Qufu 273165, China}

\author{Shu-Xiang Qiao\orcidlink{0009-0008-7092-3333}} 
\affiliation{School of Physics and Physical Engineering, Qufu Normal University, Qufu 273165, China}

\author{Pan Zhou} 
\affiliation{Hunan Provincial Key laboratory of Thin Film Materials and Devices, School of Materials Science and Engineering, Xiangtan University, Xiangtan 411105, China}

\author{Hong-Yan Lu} \email[E-mail: ]{hylu@qfnu.edu.cn}
\affiliation{School of Physics and Physical Engineering, Qufu Normal University, Qufu 273165, China}

\author{Ping Zhang}
\affiliation{School of Physics and Physical Engineering, Qufu Normal University, Qufu 273165, China}
\affiliation{Institute of Applied Physics and Computational Mathematics, Beijing 100088, China}

\begin{abstract}  
 
 Kagome superconductors featuring topologically nontrivial band structures have attracted extensive research interest. FeSn and CoSn is a new kind of kagome material with intrinsic magnetism, which suppresses the emergence of superconductivity. Here, we theoretically predict a new kind of 1:1 kagome $M$Sn ($M$=transition metal), which exhibit intrinsic superconductivity and nontrivial band topology by first-principles calculations. Among twenty-seven candidates, $M$Sn ($M$= Mo, Hf, Nb, Ta, W, Ti) are theoretically identified as both dynamically and thermodynamically stable. And, five non-magnetic $M$Sn ($M$= Mo, Hf, Nb, Ta, W) exhibit phonon-mediated superconductivity. Especially, the $d$ orbitals bands display Dirac points and van Hove singularities near the Fermi level, which contribute to the emergence of topology and the electron-phonon coupling (EPC). More interestingly, MoSn, HfSn and NbSn show nontrivial topological band structure at the Fermi level. Thus, the predicted $M$Sn establish a platform integrating superconductivity and topological order.
 
\end{abstract}

\maketitle

\section{Introduction}

Kagome-lattice materials represent a frontier in condensed matter physics, as their characteristic corner-sharing triangular geometry naturally hosts flat bands \cite{Flatband1}, van Hove singularities \cite{VHS1,VHS2,VHS3}, Dirac fermions \cite{Dirac}, and topological band structures \cite{xiaopengcheng,PhysRevB-cdwandsuperconduc,ka166,mpd5}, stabilizing diverse quantum ground states including magnetism \cite{naturefe3sn2,fesn, PRBFeSn-AFM}, charge density waves \cite{2022natureFeGe, PRB-cdw-CsV3Sb5,PRB-CDWFeGe,PRL-cdwandtopological}, nontrivial topology \cite{PRL-cdwandtopological,ka166}, and superconductivity \cite{PRL-cdwandtopological,2019PRM-superconductor,2022PRBL-superconductor}. In recent years, a large number of kagome compounds have been theoretically predicted and experimentally synthesized, such as 1:1 type CoSn \cite{CoSn,PRL-CoSn}, FeSn \cite{fesn, PRBFeSn-AFM} and FeGe \cite{fege,Nano-FeGe,2024PRB-FeGe}, 1:3 type $M$B$_{3}$ ($M$= B, Be, Ca, Sr, Mn) \cite{2023PRM-MgB3,yangl,2020PRB-MnB3}, 1:5 type $M$Pd$_{5}$ ($M$ = Ca, Sr, Ba) \cite{mpd5}, 3:1 type Mn$_{3}$$X$ ($X$ = Sn, Ge, Ga) \cite{2021PRB-Mn3X} and Ni3$_{3}$Sn \cite{Ni3Sn}, 1:3:5 type $A$V$_{3}$Sb$_{5}$ ($A$ = K, Rb,Cs) \cite{PhysRevB-cdwandsuperconduc, PRB-cdw-CsV3Sb5, 2019PRM-superconductor, 2022PRBL-superconductor, 2020PRL-CsV3Sb5}. The ground of these systems exhibit highly distinct ground state properties.

Most kagome materials exhibit rich magnetic properties. For instance, FeSn is an antiferromagnetic metal                             \cite{PRBFeSn-AFM}, Mn$_{3}$$X$ ($X$ = Sn, Ge, Ga) possesses a breathing kagome lattice with noncollinear antiferromagnetic order \cite{2021PRB-Mn3X}, and Co$_{3}$Sn$_{2}$S$_{2}$ is a ferromagnetic Weyl semimetal \cite{science-Co3Sn2S2}. Intrinsic superconductivity has been discovered in many kagome materials, including MgB$_{3}$ \cite{2023PRM-MgB3}, $M$Pd$_{5}$ ($M$ = Ca, Sr, Ba) \cite{mpd5}, and YT$_{6}$Sn$_{6}$ \cite{ka166}. The coexistence of intrinsic phonon-mediated superconductivity and nontrivial topology in $M$Pd$_{5}$ offering a new avenue to investigate the fundamental physics of topological superconductivity in kagome systems. Especially, in the kagome superconductor CsV$_{3}$Sb$_{5}$, anomalous Hall effect appears concurrently with the the higher-temperature charge density wave transition, making this system an ideal platform for studying the interplay among nontrivial topology, charge-density-wave, and superconductivity \cite{Yu2021PRB}. The coexistence of nontrivial topology and superconductivity in these materials makes them promising candidates for exploring topological superconductivity.

Recently, 1:1 type kagome materials, characterized by their simple structure, high symmetry, and clear physical picture, serve as prototypical platforms for exploring the intrinsic electronic states of kagome lattices. However, all available studies show that conventional 1:1 type kagome materials are only paramagnetic \cite{CoSn} or antiferromagnetic metals \cite{PRBFeSn-AFM}, lacking both intrinsic superconductivity.The coexistence of topology and superconductivity has never been achieved in 1:1 type kagome lattices, representing a critical gap limiting the development of this field. Although 1:3, 1:5, 1:3:5, and other families have realized the coexistence of superconductivity and topology properties \cite{2023PRM-MgB3, mpd5, 2022PRBL-superconductor}, the missing quantum states in ideal 1:1 type kagome systems prevent a complete understanding of the intrinsic correlations among electron–phonon coupling, topological and superconductivity, and hinder the establishment of universal design principles for kagome quantum materials.

In this Letter, we predict that 1:1 kagome materials, $M$Sn ($M$ = transition metals), exhibit both intrinsic superconductivity and nontrivial topological features. Among the twenty-seven 1:1 kagome $M$Sn candidates, five are non-magnetic and thermodynamically as well as dynamically stable, all exhibiting phonon-mediated superconductivity. The $d$-orbital bands display Dirac points and VHSs near the Fermi level, which contribute to the emergence of topology and the electron-phonon coupling (EPC). Moreover, three of them satisfy the criteria for topological metals. Therefore, these pristine $M$Sn materials integrate Cooper pairing with nontrivial surface states, thereby eliminating the need for external doping or heterostructure engineering. These findings not only enrich the properties of the kagome family but also establish 1:1 kagome lattice materials as a promising platform for exploring the interplay between topology and superconductivity.

\section{Computational methods}

The structural relaxation and electronic property calculations are performed within the framework of density functional theory (DFT) using the Vienna ab-initio Simulation Package (VASP) \cite{1} and the QUANTUM ESPRESSO (QE) package \cite{2}. Phonon spectra and EPC calculations are carried out using density functional perturbation theory (DFPT) \cite{66,67} as implemented in QE. The exchange-correlation functional is treated using the generalized gradient approximation (GGA) with the Perdew-Burke-Ernzerhof (PBE) parametrization \cite{3}. Electron-ion interactions are modeled using the projector augmented-wave (PAW) approach \cite{4}. In the calculation, the wave function and charge density cutoff energy are 80 and 800 Ry, respectively. 6$\times$6$\times$8 $k$-point grid is used for charge self consistent calculation, and a 3$\times$3$\times$4 $q$-point grid for dynamical matrix. For DOS calculations, 18$\times$18$\times$24 $k$-point grid is used. The surface states are obtained using the iterative Green’s function approach, implemented in the WANNIERTOOLS package \cite{5,6}, based on maximally localized Wannier functions (MLWFs) \cite{7,8} generated via the VASP2WANNIER90 interface \cite{9}. Additional computational details are provided in the Supplemental Material (SM) \cite{SM} (see also references \cite{PhysRev.167.331,DYNES1972615, PhysRevB.12.905} therein).

\section{Results and discussions}

\subsection{Crystal structure and stability}

\begin{figure}
	\centering
	\includegraphics[width=9cm]{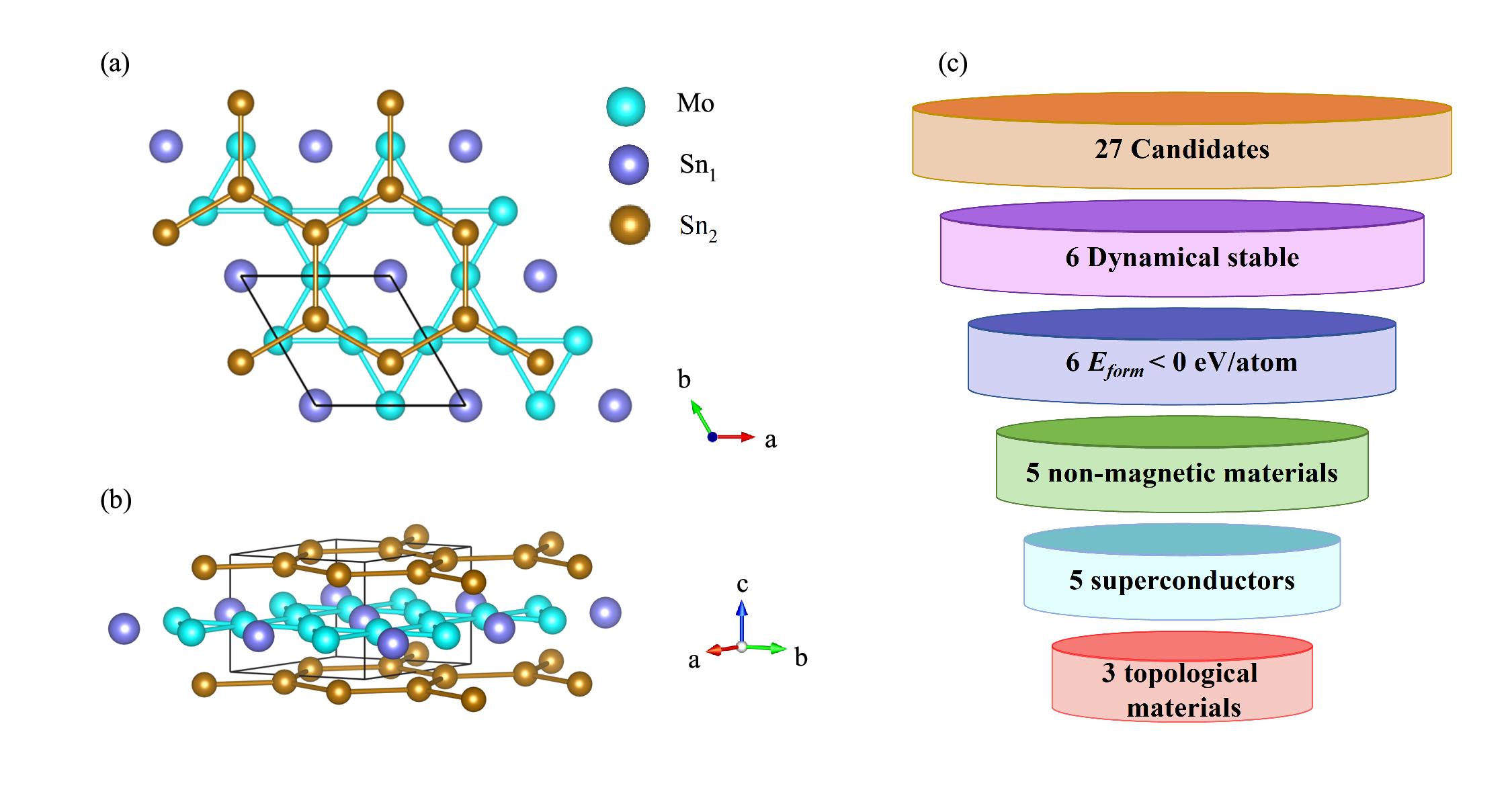}
	\caption{Top (a) and side (b) views of the crystal structure of $M$Sn ($M$=transition metals). (c) Schematic diagram of the screening process of $M$Sn. }
	\label{1}
\end{figure}

The structure of $M$Sn ($M$=transition metal)is based on the experimentally synthesized FeSn \cite{fesn}. Figure \ref{1}(a) shows the crystal structure of $M$Sn, where the $M$ (transition-metal) atoms form the kagome lattice. The primitive cell contains equal numbers of transition-metal $M$ and Sn atoms with 1:1 ratio. The Sn atoms occupy two in-equivalent lattice sites, labeled Sn$_{1}$ and Sn$_{2}$ in Fig. \ref{1}(a), lying at the hexagonal center of the $M$-kagome layer and above the triangular sites, respectively, as shown in Fig. \ref{1}(b). The space group of 1:1 kagome compound $M$Sn is $P6/mmm$ (No. 191), which incorporates kagome, triangular, and hexagonal layers, which are expected to give rise to intriguing physical properties. Figure \ref{1}(c) shows the Brillouin zone and the surface Brillouin zone projection of the $M$Sn structure. As mentioned above, FeSn and CoSn exhibit magnetic ordering and cannot display superconductivity. Atomic substitution is a widely employed strategy for the design and synthesis of new materials. Therefore, we intend to substitute Fe or Co atoms with other transition metal elements to design and identify potential superconducting and topological materials. 
\begin{table}
	\caption{The lattice constants $a$ ($\textmd{\AA}$), $c$ ($\textmd{\AA}$), and formation energy $E_{form}$ (eV/atom) of stable $M$Sn ($M$= Mo, Hf, Nb, Ta, W, Ti).}
	\centering
	\setlength{\tabcolsep}{6pt}
	\renewcommand{\arraystretch}{1}
	\begin{tabular}{cccc}
		\hline
		\hline
		Materials      & $a$    & $c$  & $E_{form}$ \\ \hline
		MoSn	       &5.61	&4.72	&-0.107     \\
		HfSn	       &6.04	&4.97	&-0.138     \\
		NbSn	       &5.80	&4.84	&-0.076     \\
		TaSn	       &5.78	&4.84	&-0.079     \\
		WSn	           &5.61	&4.75	&-0.293     \\
		TiSn 	       &5.71	&4.81	&-0.244     \\
		\hline \hline
	\end{tabular}
	\label{structure}
\end{table}
\begin{figure*}
	\centering
	\includegraphics[width=9cm]{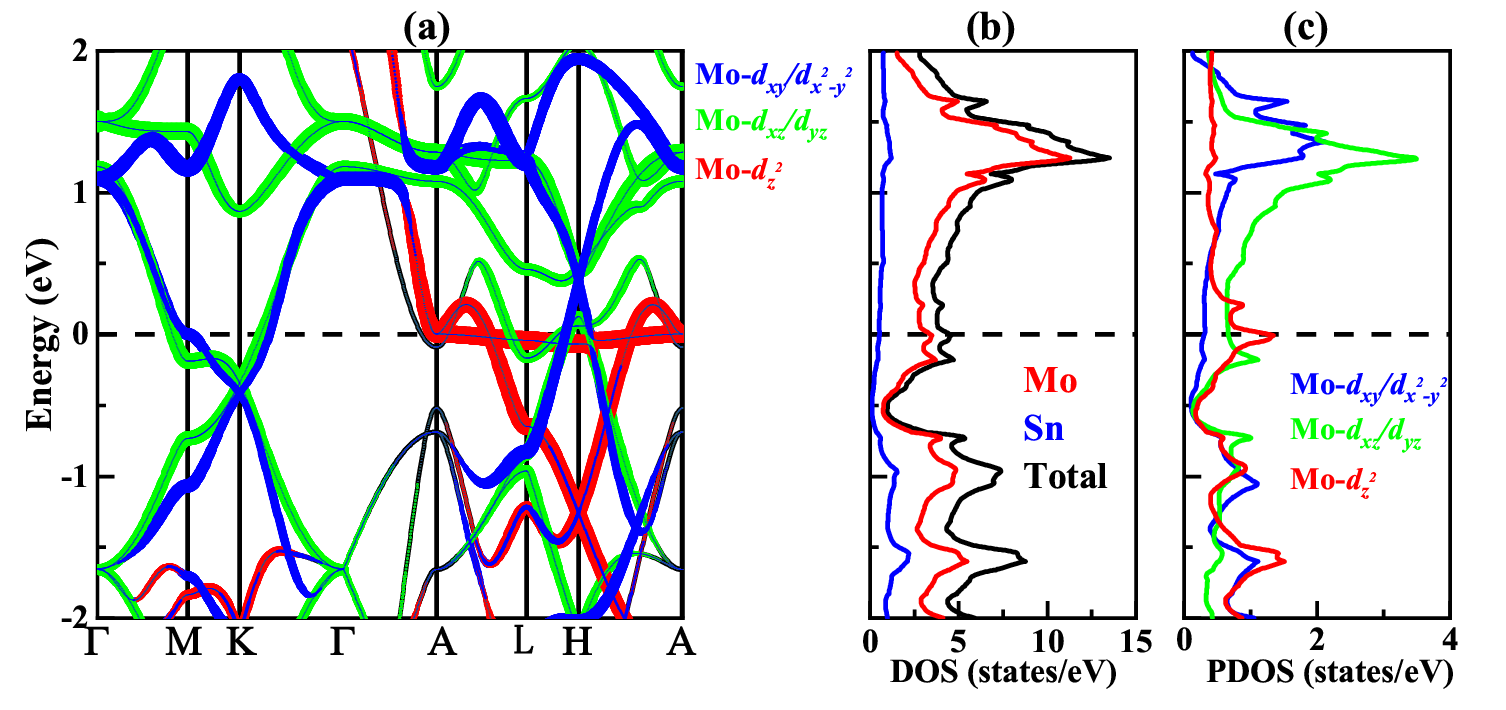}\includegraphics[width=9cm]{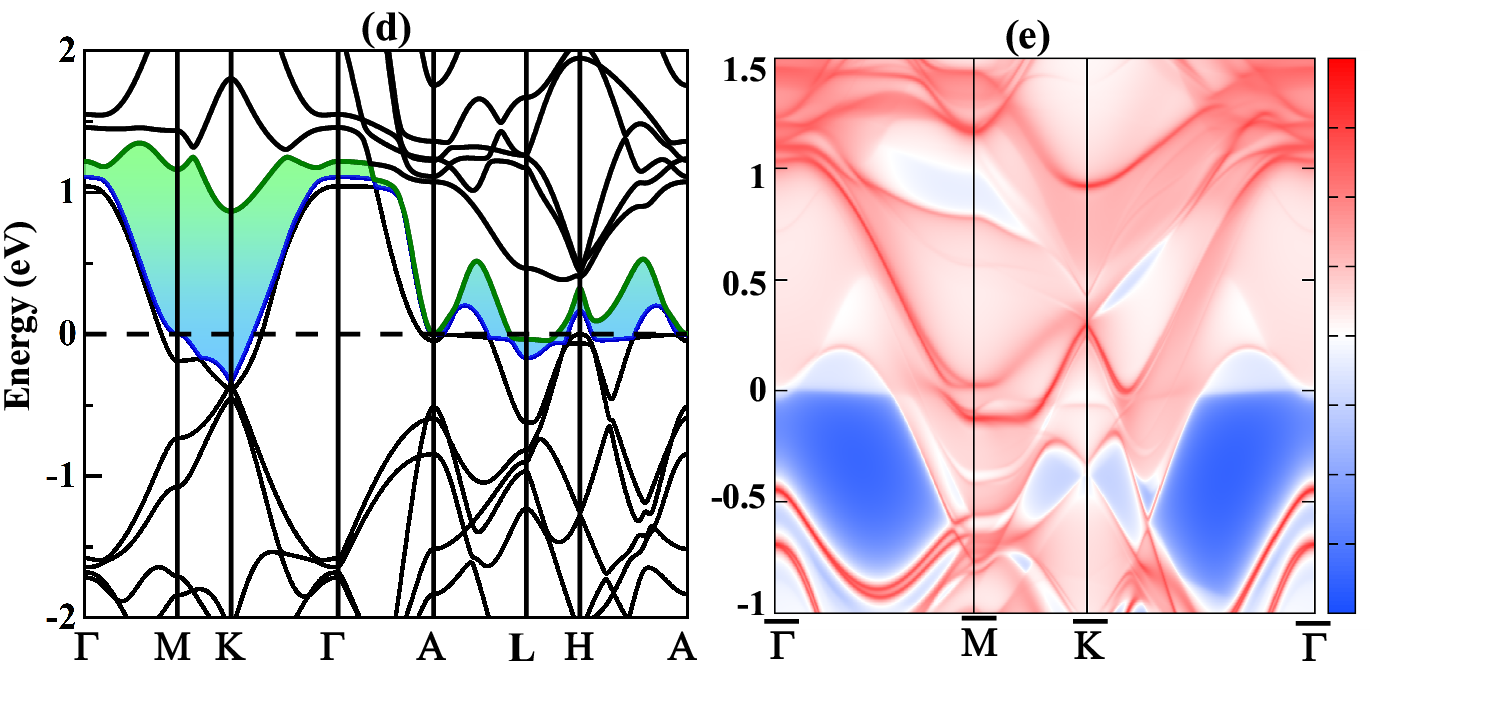}
	\includegraphics[width=9cm]{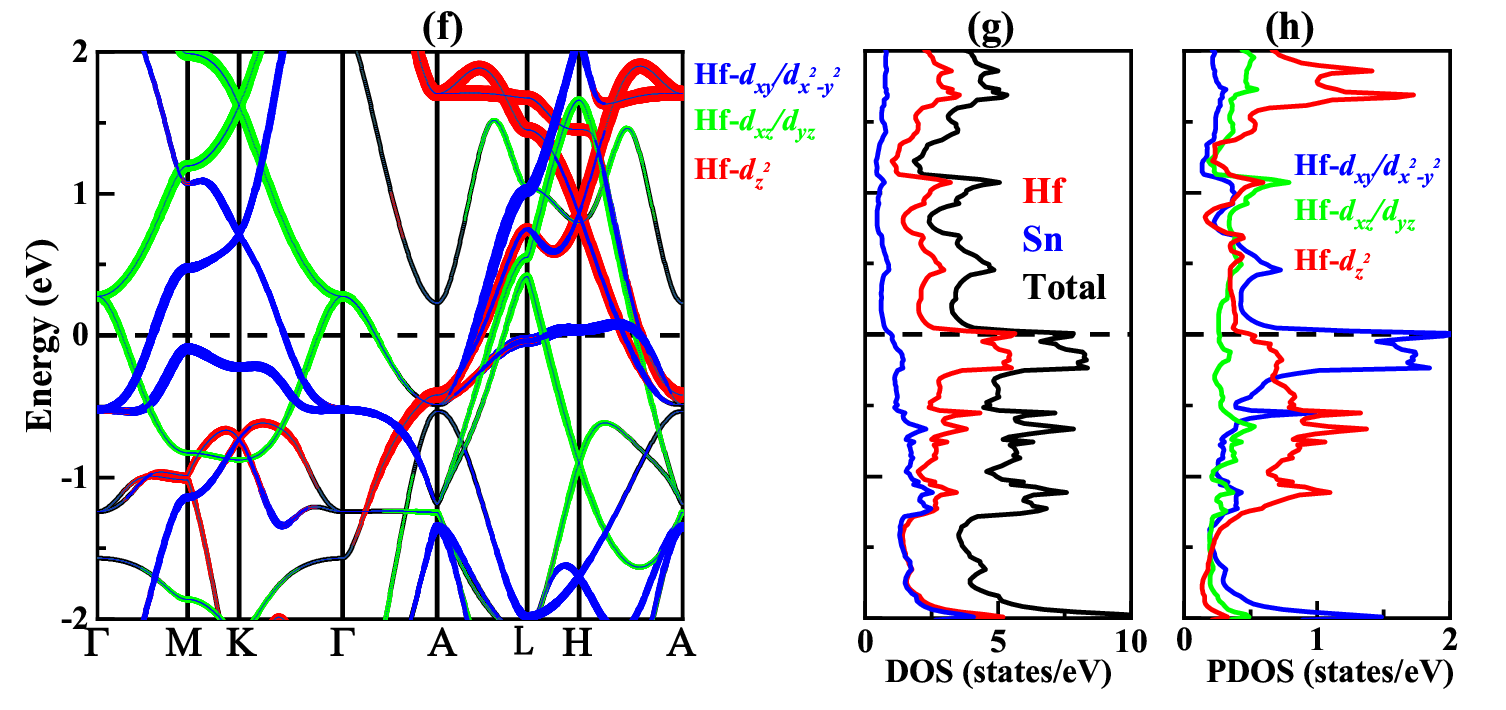}\includegraphics[width=9cm]{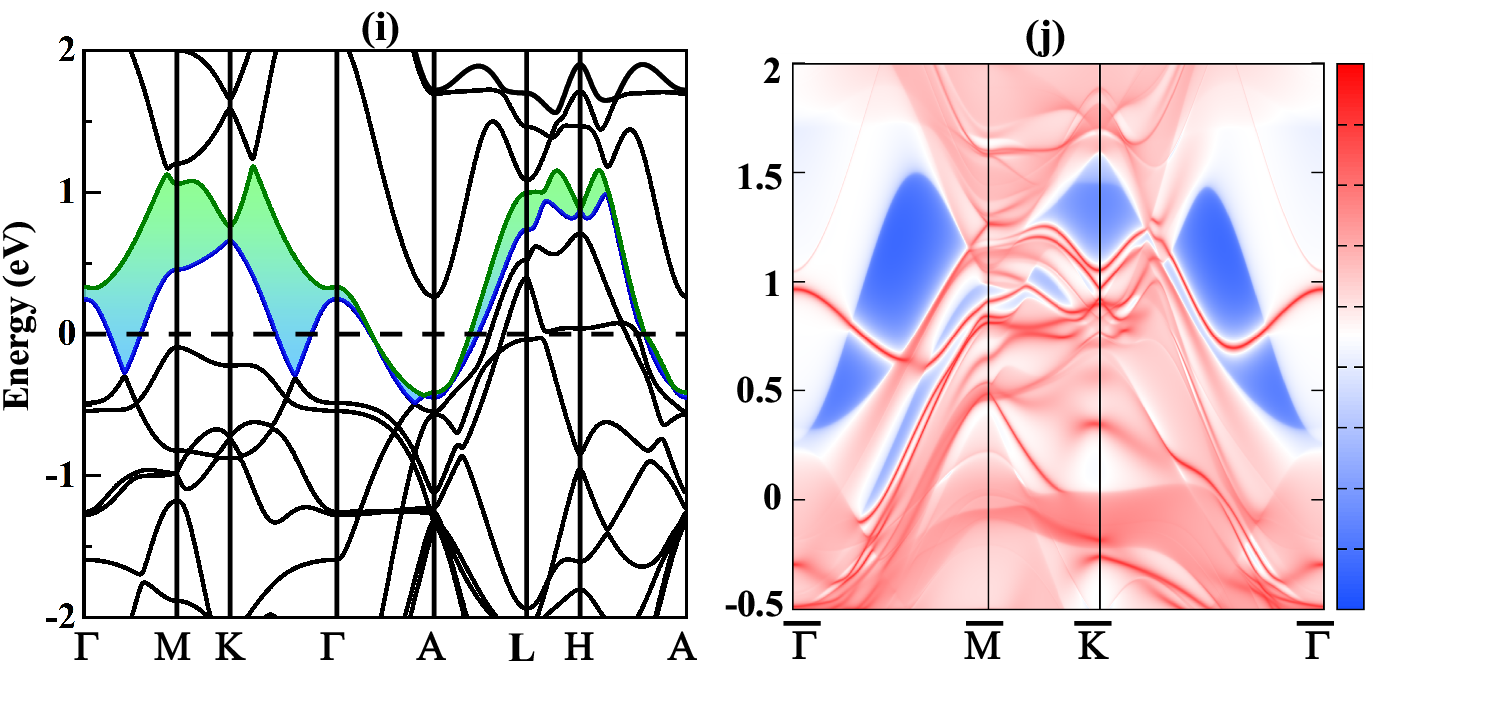}
	\caption{ (a) Orbital-projected electronic band structure without SOC, (b) total DOS, (c) PDOS, (d) band structure with SOC, and (e) surface states of MoSn. (f) to (j) are similar to (a) to (e), but for HfSn.}
	\label{2}
\end{figure*}
Here, we systematically constructed twenty-seven $M$Sn structures by substituting Fe or Co with other transition metal elements. As shown in the flowchart in Fig. \ref{1}(c), the dynamic and thermodynamic stability of the $M$Sn structures are verified using phonon spectrum and formation energy ($E_{form}$) calculations, respectively. The formation energy is defined as $E_{form}$ = [$E_{MSn}$ $-$ (3$E_{M}$+3$E_{Sn}$)]/6, where $E_{MSn}$, $E_{M}$ and $E_{Sn}$ denote the total energies of 1:1 kagome materials $M$Sn structure, the simple substance of $M$, and the Sn crystal (space group Im3m), respectively. The absence of imaginary phonon frequencies, as shown in Fig. S1 \cite{SM}, demonstrates the dynamic stability, while the negative formation energies listed in Table \ref{structure} indicate the thermodynamic stability of these structures.  Consequently, $M$Sn ($M$= Mo, Hf, Nb, Ta, W, Ti) are dynamically and thermodynamically stable, among which TiSn is found to exhibit magnetic properties and $M$Sn ($M$= Mo, Hf, Nb, Ta, W) possess non-magnetic metal properties. Therefore, the following sections mainly focus on investigating the electronic properties of these non-magnetic metal materials, including their superconductivity and electronic topology. These results are summarized in Fig. \ref{1}(c). The optimized lattice parameters ($a$ and $c$) and formation energies ($E_{form}$) of the six stable $M$Sn ($M$= Mo, Hf, Nb, Ta, W, Ti) are also listed in Table \ref{structure}, which may serve as a reference for future experimental synthesis and characterization. The experimentally realization of the predicted $M$Sn maybe can prepare via high-temperature solution growth methods, analogous to those successfully employed in the synthesis of related compounds such as FeSn\cite{fesn} and FeGe \cite{Teng2023Magnetism}. Given the structural and chemical similarities among these systems—particularly their formation through metal-rich flux environments—the self-flux method presents a feasible and straightforward route for obtaining single crystals of $M$Sn. In this approach, Sn can serve as both the reactive component and the solvent (flux), facilitating the controlled crystallization of the target phase under equilibrium conditions. This class of synthesis strategies has proven effective for kagome metallic.

\subsection{Electronic and topological properties}

To understand the novel properties of kagome materials, analyzing their electronic structures is essential. Figure \ref{2} presents the calculated electronic structures and band topology of MoSn and HfSn. The band structures of MoSn and HfSn exhibit Dirac points at the $K$ point, VHSs at the $M$ point, and flat-band-like bands along the $A$-$L$-$H$-$A$ path near the Fermi level. As can be seen from Figs. \ref{2}(b) and \ref{2}(g), the density of states near the Fermi level is primarily contributed by $M$ atoms, with a very small contribution from Sn atoms. Therefore, the weak interlayer coupling results in the DOS near the Fermi level being predominantly contributed by $M$ atoms, and the $d$-orbitals of $M$ atoms exhibit localization near the Fermi level. According to the projected band structures and orbital projected densities of states (PDOS) shown in Figs. \ref{2}(a)-\ref{2}(c) and \ref{2}(f)-\ref{2}(h), the flat-band-like bands in MoSn originate mainly from Mo-$d_{z^2}$ orbitals, while those in HfSn are primarily derived from Hf-$d_{xy}$/$d_{x^2-y^2}$ orbitals. As shown in Figs. \ref{2}(a) and \ref{2}(f), both MoSn and HfSn exhibit Dirac points at the $K$ point, which are mainly contributed by Mo-$d_{xz}$/$d_{yz}$ and Hf-$d_{xy}$/$d_{x^2-y^2}$ orbitals, respectively. The four VHSs (saddle points) at the $M$ point are also primarily derived from these two types of orbitals.  The energy bands, total DOS, and PDOS of the other three superconducting materials are shown in Fig. S2 \cite{SM}, all of which exhibit metallic behavior. Compared with the other $M$Sn materials, the saddle-point VHSs of MoSn and HfSn lie much closer to the Fermi level, leading to an enhanced DOS that may promote superconductivity \cite{vhs}.

\begin{figure*}
	\centering
	\includegraphics[width=16cm]{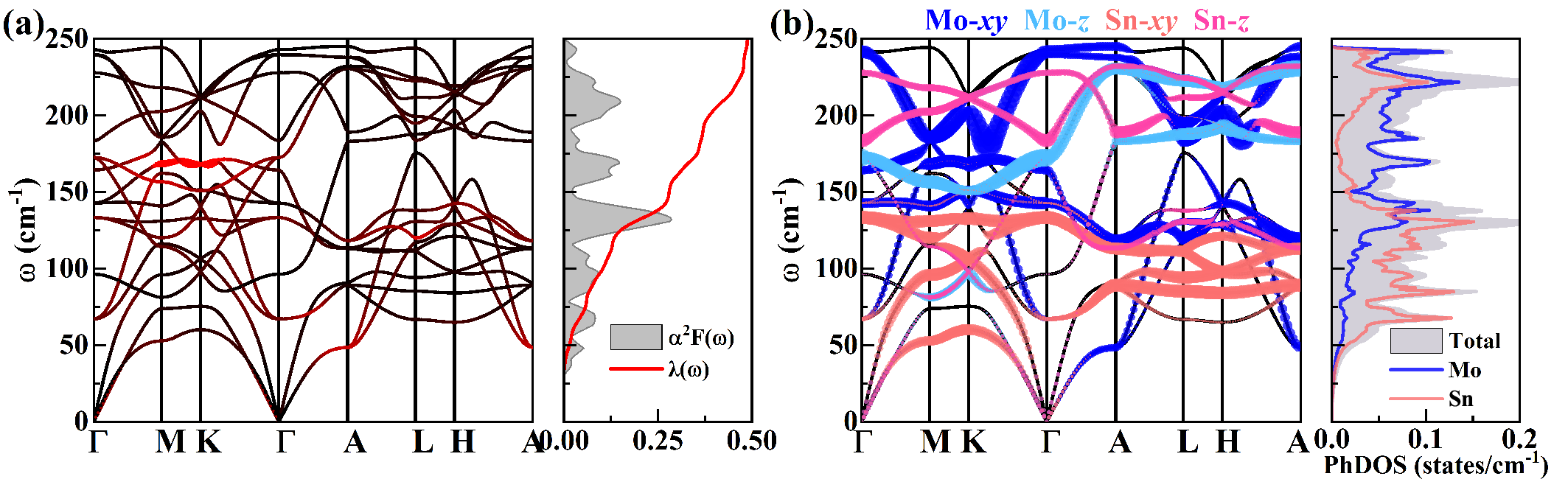}
	\includegraphics[width=16cm]{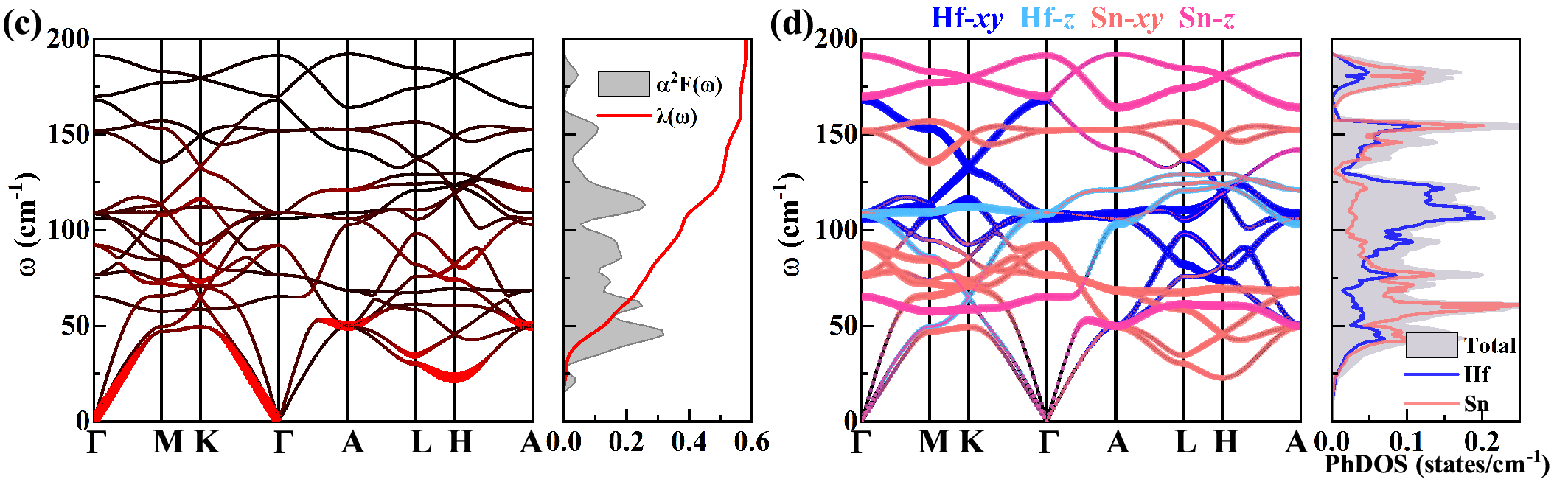}
	\caption{ (a) Phonon dispersion weighted by the magnitude of $\lambda_{\textbf{q}\nu}$ (EPC for phonon mode ${\textbf{q}\nu}$), liashberg spectral function $\alpha^{2}F (\omega)$, and EPC $\lambda (\omega)$, (b) phonon dispersion weighted by the vibration modes of each atom and PhDOS for VSn. (c) to (d) are similar to (a) to (b), but for HfSn.}
	\label{3}
\end{figure*}

The possible topological characteristics of the superconducting $M$Sn are analyzed via calculations of the $\mathbb{Z}_{2}$ topological invariant, Wannier charge centers (WCCs), and surface states \cite{PRL-cdwandtopological}. Our calculations reveal that three of the five superconducting materials (MoSn, HfSn, and NbSn) possess nontrivial topological characteristics. The electronic band structures with SOC for these three materials are shown in Figs. \ref{2}(d), \ref{2}(i), and Fig. S3 \cite{SM}. The bands near the Fermi level form continuous energy gaps throughout the Brillouin zone upon inclusion of SOC, as highlighted in Figs. \ref{2}(d), \ref{2}(i), and Fig. S3 \cite{SM}. The continuous energy gaps enable the determination of the $\mathbb{Z}_{2}$ topological invariant, defined in the same manner as in Bi$_{2}$Se$_{3}$ \cite{Bi2Se3}.

The $\mathbb{Z}_{2}$ topological invariant is subsequently calculated using the WCC method. For these three materials, an arbitrary horizontal reference line intersects the WCC evolution curves an odd number of times, indicating the $\mathbb{Z}_{2}$ value of 1, as shown in Fig. S4 \cite{SM}. Similar to other kagome metals such as $YT_{6}$Sn$_{5}$ \cite{ka166} and $M$Pd$_{5}$ \cite{mpd5}, MoSn, HfSn, and NbSn preserve both spatial inversion and time-reversal symmetries. Figures \ref{2}(e), \ref{2}(j), and Fig. S3 illustrate the (001) surface states of MoSn, HfSn, and NbSn, and it is found that there are prominent surface states near the Fermi level. These surface states clearly cross the Fermi level, suggesting potential avenues for realizing topological superconductivity. The presence of nontrivial topological surface states within the bulk gap further supports the nontrivial band topology of these materials. Therefore, MoSn, HfSn, and NbSn exhibit nontrivial topological band properties and can be classified as topological metals.

\subsection{EPC and superconductivity}

Given that $M$Sn ($M$= Mo, Hf, Nb, Ta, W) exhibit non-magnetic metallic properties and flat band characteristics, we further explore their potential for superconductivity. As mentioned in the previous section, the flat bands of MoSn and HfSn are located at the Fermi level, resulting in high DOS, which enhances the EPC and promotes stronger superconductivity. Furthermore, we calculated the phonon dispersion weighted by the magnitude of EPC linewidth $\lambda_{\textbf{q}\nu}$, Eliashberg spectral function $\alpha^{2}F (\omega)$, EPC $\lambda (\omega)$, phonon dispersion weighted by the vibration modes contributed by different elements, and phonon DOS (PhDOS) of MoSn and HfSn, as shown in Fig. \ref{3}. For kagome materials $M$Sn, the primitive cell contains six atoms, giving rise to eighteen phonon branches, including three acoustic and fifteen optical modes.

To analyze the phonon mode contributions to the superconducting mechanism in MoSn and HfSn, we compare the phonon dispersions associated with different atomic vibration modes (Figs. \ref{3}(a), \ref{3}(c)) and the distribution and strength of the EPC (Figs. \ref{3}(b), \ref{3}(d)). For MoSn, the vibration modes near the $M$ and $K$ points (around 170 cm$^{-1}$) exhibit significant EPC effects, resulting in a pronounced enhancement of the EPC $\lambda$. The intermediate-frequency phonon modes (130-170 cm$^{-1}$) contribute approximately 56 $\%$ to the total EPC constant $\lambda$. As shown in Fig. \ref{3}(b), due to the difference in atomic masses, the atomic vibration frequencies exhibit a relatively separation. The vibrations of Mo atoms are primarily concentrated in the high-frequency region, whereas the vibrations of Sn atoms dominate the low-frequency region. Through comparison, the dominant EPC contributions arise mainly from the Mo-$xy$ plane vibration modes. Vibrations in other frequency ranges are also distributed throughout the spectrum and make additional contributions to the EPC.

For HfSn, as shown in Fig. \ref{3}(c), the EPC mainly originates from low-frequency vibrations (0-100 cm$^{-1}$), which contribute approximately 71 $\%$ to the total EPC constant $\lambda$. The vibrational frequency distributions of Hf and Sn atoms are shown in Fig. \ref{3}(d). It is observed that the vibrations of Sn atoms are mainly concentrated in the low-frequency region, and the phonon mode dominated by the Sn-$xy$ plane vibrations shows pronounced softening near the $H$ point, which strongly contributes to the enhanced EPC strength. Similar to MoSn, vibrations in other frequency ranges also contribute to the EPC, though to a lesser extent. The EPC calculation results for the other three superconducting $M$Sn are shown in Fig. S4 \cite{SM}.

\begin{figure}
	\centering
	\includegraphics[width=8cm]{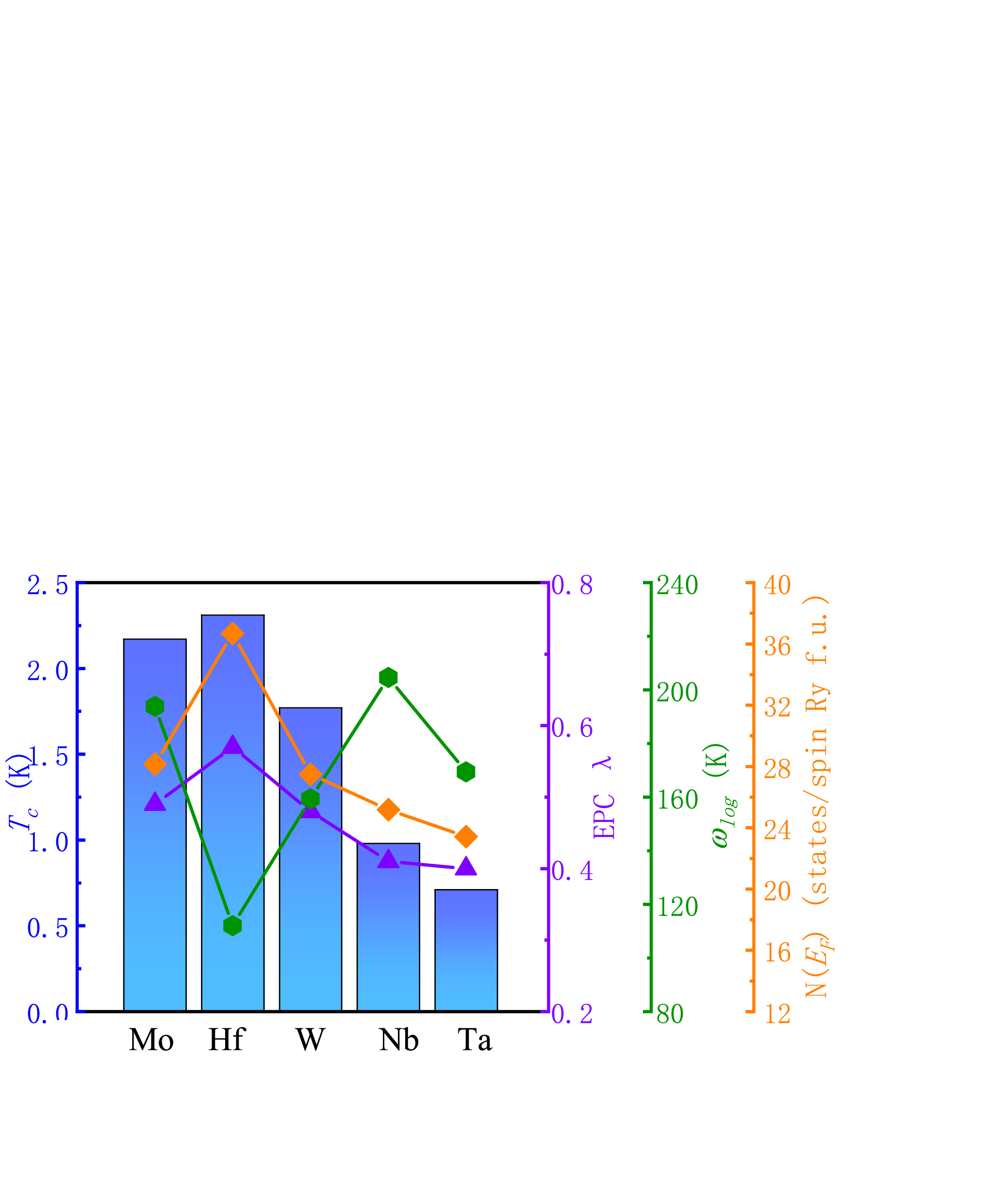}
	\caption{ Variation of $T_{c}$, EPC $\lambda$, $\omega_{log}$, and $N(E_{F})$ for superconducting $M$Sn.}
	\label{4}
\end{figure}

According to our calculations, the total EPC constants $\lambda$ for MoSn and HfSn are 0.49 and 0.57, respectively, indicating that both materials can be categorized as weak coupling superconductors. The calculated superconducting $T_{c}$ are 2.17 K for MoSn and 2.31 K for HfSn. In addition, the corresponding data for the other three superconducting $M$Sn are summarized in Fig. \ref{4} and Table S1 \cite{SM}. It is evident that the $T_{c}$ values of the 1:1 kagome materials are positively correlated with the EPC $\lambda$ and the DOS at the Fermi level $N(E_{F})$, suggesting that they are conventional phonon-mediated superconductors. Therefore, our results provide valuable insights and opportunities for exploring the novel physical properties of 1:1 kagome materials.

Importantly, our calculations reveal that MoSn, HfSn, and NbSn are superconducting and topological metals (SCTMs) with coexisting superconductivity and nontrivial topological properties. Experimentally, superconductivity in topological materials is typically induced by doping topological insulators (such as Cu- or Nb-doped Bi$_{2}$Se$_{3}$ \cite{CuBiSe,NbBiSe} and Sn$_{1-x}$In$_{x}$Te \cite{SnTe}) or by proximity effects in hetero-structures \cite{tsc1,Hu2025Numerical}. Unlike the charge-density-wave (CDW) dominated AV$_{3}$Sb$_{5}$ (A = K, Rb, Cs) family, where superconductivity emerges under pressure\cite{PhysRevB-cdwandsuperconduc}, $M$Sn exhibits intrinsic coexistence of topological band features and superconductivity. Furthermore, while FeSn \cite{fesn} and FeGe \cite{fege} share a similar kagome crystal but display itinerant magnetism rather than bulk superconductivity. Similar with the recently discovered kagome superconductor YT$_{6}$Sn$_{6}$ (T = V, Rb, Ta) \cite{ka166}, $M$Sn ($M$= Mo, Hf, Nb) achieves a rare combination of superconductivity and topological robustness without magnetic ordering. These findings may serve as valuable references for future experimental and theoretical studies. 

\section{Conclusion}

In summary, we have predicted a new class of 1:1 kagome materials, unveiling their structural stability, topological, and superconducting properties. Through comprehensive screening of twenty-seven candidate materials, we identified $M$Sn ($M$= Mo, Hf, Nb, Ta, W, Ti) are dynamically and thermally stable. $M$Sn ($M$= Mo, Hf, Nb, Ta, W) were further confirmed to be superconductors and the $T_{c}$ ranged from 0.71 K to 2.31 K. It is worth noting that the flat-band-like and the VHSs of MoSn and HfSn are near the Fermi level, resulting in high DOS. More interestingly, MoSn, HfSn, and NbSn are identified as SCTMs, whose surface states cross the Fermi level. Thus, the predicted $M$Sn establish a multifunctional platform integrating superconductivity and topological order without the need for external doping or hetero-structure fabrication. 

\section{Acknowledgements}
 This work is supported by the National Natural Science Foundation of China (Grant No. 12074213), the National Key R$\&$D Program of China (Grant No. 2022YFA1403103), the Major Basic Program of Natural Science Foundation of Shandong Province (Grant No. ZR2021ZD01), and the Natural Science Foundation of Shandong Province (Grant No. ZR2023MA082).

\bibliography{references}

@article{PhysRev.167.331,
  title = {Transition Temperature of Strong-Coupled Superconductors},
  author = {McMillan, W. L.},
  journal = {Phys. Rev.},
  volume = {167},
  issue = {2},
  pages = {331--344},
  numpages = {0},
  year = {1968},
  month = {Mar},
  publisher = {American Physical Society},
  doi = {10.1103/PhysRev.167.331},
  url = {https://link.aps.org/doi/10.1103/PhysRev.167.331}
}

@article{DYNES1972615,
title = {McMillan's equation and the Tc of superconductors},
journal = {Solid State Communications},
volume = {10},
number = {7},
pages = {615-618},
year = {1972},
issn = {0038-1098},
doi = {https://doi.org/10.1016/0038-1098(72)90603-5},
url = {https://www.sciencedirect.com/science/article/pii/0038109872906035},
author = {R.C. Dynes}
}

@article{PhysRevB.12.905,
  title = {Transition temperature of strong-coupled superconductors reanalyzed},
  author = {Allen, P. B. and Dynes, R. C.},
  journal = {Phys. Rev. B},
  volume = {12},
  issue = {3},
  pages = {905--922},
  numpages = {0},
  year = {1975},
  month = {Aug},
  publisher = {American Physical Society},
  doi = {10.1103/PhysRevB.12.905},
  url = {https://link.aps.org/doi/10.1103/PhysRevB.12.905}
}

@article{2022NatureFeGe,
  title     = {Discovery of charge density wave in a kagome lattice antiferromagnet},
  author    = {Teng, Xiaokun and Chen, Lebing and Ye, Feng and Rosenberg, Elliott and Liu, Zhaoyu and Yin, Jia Xin and Jiang, Yu Xiao and Oh, Ji Seop and Hasan, M. Zahid and Neubauer, Kelly J. and Gao, Bin and Xie, Yaofeng and Hashimoto, Makoto and Lu, Donghui and Jozwiak, Chris and Bostwick, Aaron and Rotenberg, Eli and Birgeneau, Robert J. and Chu, Jiun Haw and Yi, Ming and Dai, Pengcheng},
  journal   = {Nature},
  year      = {2022},
  volume    = {609},
  number    = {7927},
  pages     = {490--495},
  doi       = {10.1038/s41586-022-05034-z},
  issn      = {1476-4687},
  url       = {https://doi.org/10.1038/s41586-022-05034-z}
}

@article{Teng2023Magnetism,
  title     = {Magnetism and charge density wave order in kagome {F}e{G}e},
  author    = {Teng, Xiaokun and Oh, Ji Seop and Tan, Hengxin and Chen, Lebing and Huang, Jianwei and Gao, Bin and Yin, Jia Xin and Chu, Jiun Haw and Hashimoto, Makoto and Lu, Donghui and Jozwiak, Chris and Bostwick, Aaron and Rotenberg, Eli and Granroth, Garrett E. and Yan, Binghai and Birgeneau, Robert J. and Dai, Pengcheng and Yi, Ming},
  journal   = {Nature Physics},
  year      = {2023},
  volume    = {19},
  number    = {6},
  pages     = {814--822},
  doi       = {10.1038/s41567-023-01985-w},
  issn      = {1745-2481},
  url       = {https://doi.org/10.1038/s41567-023-01985-w}
}

@article{Yu2021PRB,
  title = {Concurrence of anomalous Hall effect and charge density wave in a superconducting topological kagome metal},
  author = {Yu, F. H. and Wu, T. and Wang, Z. Y. and Lei, B. and Zhuo, W. Z. and Ying, J. J. and Chen, X. H.},
  journal = {Phys. Rev. B},
  volume = {104},
  issue = {4},
  pages = {L041103},
  numpages = {7},
  year = {2021},
  month = {Jul},
  publisher = {American Physical Society},
  doi = {10.1103/PhysRevB.104.L041103},
  url = {https://link.aps.org/doi/10.1103/PhysRevB.104.L041103}
}

@article{PRB-CDWFeGe,
  title = {Interacting spin and charge density waves in the kagome metal {F}e{G}e},
  author = {Klemm, Mason L. and Zhang, Tingjun and Winn, Barry L. and Li, Fankang and Ye, Feng and Matsuda, Masaaki and Maity, Avishek and Xu, Sijie and Teng, Xiaokun and Umemoto, Yoshihiko and Gao, Bin and Yi, Ming and Dai, Pengcheng},
  journal = {Phys. Rev. B},
  volume = {112},
  issue = {17},
  pages = {174422},
  numpages = {9},
  year = {2025},
  month = {Nov},
  publisher = {American Physical Society},
  doi = {10.1103/d88l-4wq8},
  url = {https://link.aps.org/doi/10.1103/d88l-4wq8}
}

@article{PRB-cdw-CsV3Sb5,
  title = {Coherent phonon pairs and rotational symmetry breaking of charge density wave order in the kagome superconductor {C}s{V}$_{3}${Sb}$_{5}$},
  author = {Deng, Qinwen and Tan, Hengxin and Ortiz, Brenden R. and Salinas, Andrea Capa and Wilson, Stephen D. and Yan, Binghai and Wu, Liang},
  journal = {Phys. Rev. B},
  volume = {112},
  issue = {12},
  pages = {125127},
  numpages = {13},
  year = {2025},
  month = {Sep},
  publisher = {American Physical Society},
  doi = {10.1103/gmdl-qz2t},
  url = {https://link.aps.org/doi/10.1103/gmdl-qz2t}
}

@Misc{SM,
  title = {See {Supplemental} {Material} for calculation details, phonon spectra of all {$M$}{S}n, band, {DOS}, band topology, and superconductivity for superconducting {$M$}{S}n.},
}

@Article{vhs,
  author    = {Kim, Kyoo and Kim, Sooran and Kim, J. S. and Kim, Heejung and Park, J.-H. and Min, B. I.},
  journal   = {Phys. Rev. B},
  title     = {Importance of the van Hove singularity in superconducting {P}d{T}e$_{2}$},
  year      = {2018},
  month     = {Apr},
  pages     = {165102},
  volume    = {97},
  doi       = {10.1103/PhysRevB.97.165102},
  issue     = {16},
  numpages  = {5},
  publisher = {American Physical Society},
  url       = {https://link.aps.org/doi/10.1103/PhysRevB.97.165102},
}

@Article{ka166,
  author    = {Shi, Lan-Ting and Si, Jian-Guo and Liang, Akun and Turnbull, Robin and Liu, Peng-Fei and Wang, Bao-Tian},
  journal   = {Phys. Rev. B},
  title     = {Topological and superconducting properties in bilayer kagome metals {Y}{T}$_{6}${S}n$_{6}$ ({T}={V}, {N}b, {T}a)},
  year      = {2023},
  month     = {May},
  pages     = {184503},
  volume    = {107},
  doi       = {10.1103/PhysRevB.107.184503},
  issue     = {18},
  numpages  = {9},
  publisher = {American Physical Society},
  url       = {https://link.aps.org/doi/10.1103/PhysRevB.107.184503},
}

@Article{PRL-cdwandtopological,
  author    = {Tan, Hengxin and Liu, Yizhou and Wang, Ziqiang and Yan, Binghai},
  journal   = {Phys. Rev. Lett.},
  title     = {Charge Density Waves and Electronic Properties of Superconducting Kagome Metals},
  year      = {2021},
  month     = {Jul},
  pages     = {046401},
  volume    = {127},
  doi       = {10.1103/PhysRevLett.127.046401},
  issue     = {4},
  numpages  = {6},
  publisher = {American Physical Society},
  url       = {https://link.aps.org/doi/10.1103/PhysRevLett.127.046401},
}

@Article{mpd5,
  author    = {Li, Dan and Wang, Zhengxuan and Jing, Panshi and Shiri, Mehrdad and Wang, Kun and Ma, Chunlan and Gong, Shijing and Zhao, Chuanxi and Wang, Tianxing and Dong, Xiao and Zhuang, Lin and Liu, Wuming and An, Yipeng},
  journal   = {Phys. Rev. B},
  title     = {{M}{P}d$_{5}$ kagome superconductors studied by density functional calculations},
  year      = {2025},
  month     = {Apr},
  pages     = {144511},
  volume    = {111},
  doi       = {10.1103/PhysRevB.111.144511},
  issue     = {14},
  numpages  = {11},
  publisher = {American Physical Society},
  url       = {https://link.aps.org/doi/10.1103/PhysRevB.111.144511},
}

@article{science-Co3Sn2S2,
author = {D. F. Liu  and A. J. Liang  and E. K. Liu  and Q. N. Xu  and Y. W. Li  and C. Chen  and D. Pei  and W. J. Shi  and S. K. Mo  and P. Dudin  and T. Kim  and C. Cacho  and G. Li  and Y. Sun  and L. X. Yang  and Z. K. Liu  and S. S. P. Parkin  and C. Felser  and Y. L. Chen },
title = {Magnetic Weyl semimetal phase in a Kagome crystal},
journal = {Science},
volume = {365},
number = {6459},
pages = {1282-1285},
year = {2019},
doi = {10.1126/science.aav2873},
URL = {https://www.science.org/doi/abs/10.1126/science.aav2873}}

@article{2021PRB-Mn3X,
  title = {Anisotropic magnetic interactions in hexagonal $AB$-stacked kagome lattice structures: Application to {M}n$_{3}${X} ({X}={G}e,{S}n,{G}a) compounds},
  author = {Zelenskiy, A. and Monchesky, T. L. and Plumer, M. L. and Southern, B. W.},
  journal = {Phys. Rev. B},
  volume = {103},
  issue = {14},
  pages = {144401},
  numpages = {11},
  year = {2021},
  month = {Apr},
  publisher = {American Physical Society},
  doi = {10.1103/PhysRevB.103.144401},
  url = {https://link.aps.org/doi/10.1103/PhysRevB.103.144401}
}

@Article{CuBiSe,
  author    = {Hor, Y. S. and Williams, A. J. and Checkelsky, J. G. and Roushan, P. and Seo, J. and Xu, Q. and Zandbergen, H. W. and Yazdani, A. and Ong, N. P. and Cava, R. J.},
  journal   = {Phys. Rev. Lett.},
  title     = {Superconductivity in {Cu}$_{x}${Bi}$_{2}${S}e$_{3}$ and its Implications for Pairing in the Undoped Topological Insulator},
  year      = {2010},
  month     = {Feb},
  pages     = {057001},
  volume    = {104},
  doi       = {10.1103/PhysRevLett.104.057001},
  issue     = {5},
  numpages  = {4},
  publisher = {American Physical Society},
  url       = {https://link.aps.org/doi/10.1103/PhysRevLett.104.057001},
}

@Article{NbBiSe,
  author    = {Asaba, Tomoya and Lawson, B. J. and Tinsman, Colin and Chen, Lu and Corbae, Paul and Li, Gang and Qiu, Y. and Hor, Y. S. and Fu, Liang and Li, Lu},
  journal   = {Phys. Rev. X},
  title     = {Rotational Symmetry Breaking in a Trigonal Superconductor Nb-doped {Bi}$_{2}${S}e$_{3}$},
  year      = {2017},
  month     = {Jan},
  pages     = {011009},
  volume    = {7},
  doi       = {10.1103/PhysRevX.7.011009},
  issue     = {1},
  numpages  = {7},
  publisher = {American Physical Society},
  url       = {https://link.aps.org/doi/10.1103/PhysRevX.7.011009},
}

@Article{SnTe,
  author    = {Balakrishnan, G. and Bawden, L. and Cavendish, S. and Lees, M. R.},
  journal   = {Phys. Rev. B},
  title     = {Superconducting properties of the In-substituted topological crystalline insulator {S}n{T}e},
  year      = {2013},
  month     = {Apr},
  pages     = {140507},
  volume    = {87},
  doi       = {10.1103/PhysRevB.87.140507},
  issue     = {14},
  numpages  = {4},
  publisher = {American Physical Society},
  url       = {https://link.aps.org/doi/10.1103/PhysRevB.87.140507},
}

@Article{tsc1,
  author    = {Hu, Jingnan and Yu, Fei and Luo, Aiyun and Pan, Xiao Hong and Zou, Jinyu and Liu, Xin and Xu, Gang},
  journal   = {Phys. Rev. Lett.},
  title     = {Chiral Topological Superconductivity in Superconductor-Obstructed Atomic Insulator-Ferromagnetic Insulator Heterostructures},
  year      = {2024},
  month     = {Jan},
  pages     = {036601},
  volume    = {132},
  doi       = {10.1103/PhysRevLett.132.036601},
  issue     = {3},
  numpages  = {7},
  publisher = {American Physical Society},
  url       = {https://link.aps.org/doi/10.1103/PhysRevLett.132.036601},
}

@article{Hu2025Numerical,
  title     = {A numerical method for designing topological superconductivity induced by s-wave pairing},
  author    = {Hu, Jingnan and Luo, Aiyun and Wang, Zhijun and Zou, Jingyu and Wu, Quansheng and Xu, Gang},
  journal   = {npj Computational Materials},
  year      = {2025},
  volume    = {11},
  number    = {1},
  pages     = {133},
  doi       = {10.1038/s41524-025-01621-6},
  issn      = {2057-3960},
  url       = {https://doi.org/10.1038/s41524-025-01621-6}
}

@article{2019PRM-superconductor,
  title = {New kagome prototype materials: discovery of {KV}$_{3}${Sb}$_{5}$,{R}b{V}$_{3}${S}b$_{5}$, and {C}s{V}$_{3}${S}b$_{5}$},
  author = {Ortiz, Brenden R. and Gomes, L\'{\i}dia C. and Morey, Jennifer R. and Winiarski, Michal and Bordelon, Mitchell and Mangum, John S. and Oswald, Iain W. H. and Rodriguez-Rivera, Jose A. and Neilson, James R. and Wilson, Stephen D. and Ertekin, Elif and McQueen, Tyrel M. and Toberer, Eric S.},
  journal = {Phys. Rev. Mater.},
  volume = {3},
  issue = {9},
  pages = {094407},
  numpages = {9},
  year = {2019},
  month = {Sep},
  publisher = {American Physical Society},
  doi = {10.1103/PhysRevMaterials.3.094407},
  url = {https://link.aps.org/doi/10.1103/PhysRevMaterials.3.094407}
}

@article{2022PRBL-superconductor,
  title = {Gapless excitations inside the fully gapped kagome superconductors {AV}$_{3}${S}b$_{5}$},
  author = {Gu, Yuhao and Zhang, Yi and Feng, Xilin and Jiang, Kun and Hu, Jiangping},
  journal = {Phys. Rev. B},
  volume = {105},
  issue = {10},
  pages = {L100502},
  numpages = {5},
  year = {2022},
  month = {Mar},
  publisher = {American Physical Society},
  doi = {10.1103/PhysRevB.105.L100502},
  url = {https://link.aps.org/doi/10.1103/PhysRevB.105.L100502}
}

@article{2020PRL-CsV3Sb5,
  title = {{C}s{V}$_{3}${S}b$_{5}$: A {Z}$_{2}$ Topological Kagome Metal with a Superconducting Ground State},
  author = {Ortiz, Brenden R. and Teicher, Samuel M. L. and Hu, Yong and Zuo, Julia L. and Sarte, Paul M. and Schueller, Emily C. and Abeykoon, A. M. Milinda and Krogstad, Matthew J. and Rosenkranz, Stephan and Osborn, Raymond and Seshadri, Ram and Balents, Leon and He, Junfeng and Wilson, Stephen D.},
  journal = {Phys. Rev. Lett.},
  volume = {125},
  issue = {24},
  pages = {247002},
  numpages = {6},
  year = {2020},
  month = {Dec},
  publisher = {American Physical Society},
  doi = {10.1103/PhysRevLett.125.247002},
  url = {https://link.aps.org/doi/10.1103/PhysRevLett.125.247002}
}

@article{PhysRevB-cdwandsuperconduc,
  title     = {Charge density wave and pressure-dependent superconductivity in the kagome metal {C}s{V}$_{3}${S}b$_{5}$ : A first-principles study},
  author    = {Si, Jian Guo and Lu, Wen Jian and Sun, Yu Ping and Liu, Peng Fei and Wang, Bao-Tian},
  journal   = {Phys. Rev. B},
  year      = {2022},
  volume    = {105},
  number    = {2},
  pages     = {024517},
  doi       = {10.1103/PhysRevB.105.024517},
  url       = {https://link.aps.org/doi/10.1103/PhysRevB.105.024517},
  publisher = {American Physical Society}
}

@Article{1,
  author    = {Kresse, G. and Furthm\"uller, J.},
  journal   = {Phys. Rev. B},
  title     = {Efficient iterative schemes for ab initio total-energy calculations using a plane-wave basis set},
  year      = {1996},
  month     = {Oct},
  pages     = {11169--11186},
  volume    = {54},
  doi       = {10.1103/PhysRevB.54.11169},
  issue     = {16},
  numpages  = {0},
  publisher = {American Physical Society},
  url       = {https://link.aps.org/doi/10.1103/PhysRevB.54.11169},
}

@Article{2,
  author   = {Giannozzi, Paolo and Baroni, Stefano and Bonini, Nicola and Calandra, Matteo and Car, Roberto and Cavazzoni, Carlo and Ceresoli, Davide and Chiarotti, Guido L. and Cococcioni, Matteo and Dabo, Ismaila and Dal Corso, Andrea and de Gironcoli, Stefano and Fabris, Stefano and Fratesi, Guido and Gebauer, Ralph and Gerstmann, Uwe and Gougoussis, Christos and Kokalj, Anton and Lazzeri, Michele and Martin Samos, Layla and Marzari, Nicola and Mauri, Francesco and Mazzarello, Riccardo and Paolini, Stefano and Pasquarello, Alfredo and Paulatto, Lorenzo and Sbraccia, Carlo and Scandolo, Sandro and Sclauzero, Gabriele and Seitsonen, Ari P. and Smogunov, Alexander and Umari, Paolo and Wentzcovitch, Renata M.},
  journal  = {J. Phys.: Condens. Matter},
  title    = {QUANTUM ESPRESSO: a modular and open-source software project for quantum simulations of materials},
  year     = {2009},
  issn     = {0953-8984},
  number   = {39},
  pages    = {395502},
  volume   = {21},
  doi      = {10.1088/0953-8984/21/39/395502},
  url      = {https://dx.doi.org/10.1088/0953-8984/21/39/395502},
}

@Article{5,
  author   = {Wu, QuanSheng and Zhang, ShengNan and Song, Hai Feng and Troyer, Matthias and Soluyanov, Alexey A.},
  journal  = {Comput. Phys. Commun.},
  title    = {WannierTools: An open-source software package for novel topological materials},
  year     = {2018},
  issn     = {0010-4655},
  pages    = {405--416},
  volume   = {224},
  doi      = {10.1016/j.cpc.2017.09.033},
  url      = {https://www.sciencedirect.com/science/article/pii/S0010465517303442},
}

@Article{6,
  author   = {Sancho, M. P. Lopez and Sancho, J. M. Lopez and Sancho, J. M. L. and Rubio, J.},
  journal  = {J. Phys. F: Met. Phys.},
  title    = {Highly convergent schemes for the calculation of bulk and surface Green functions},
  year     = {1985},
  issn     = {0305-4608},
  number   = {4},
  pages    = {851},
  volume   = {15},
  doi      = {10.1088/0305-4608/15/4/009},
  url      = {https://dx.doi.org/10.1088/0305-4608/15/4/009},
}

@Article{7,
  author    = {Marzari, Nicola and Mostofi, Arash A. and Yates, Jonathan R. and Souza, Ivo and Vanderbilt, David},
  journal   = {Rev. Mod. Phys.},
  title     = {Maximally localized Wannier functions: Theory and applications},
  year      = {2012},
  month     = {Oct},
  pages     = {1419--1475},
  volume    = {84},
  doi       = {10.1103/RevModPhys.84.1419},
  issue     = {4},
  numpages  = {0},
  publisher = {American Physical Society},
  url       = {https://link.aps.org/doi/10.1103/RevModPhys.84.1419},
}

@Article{8,
  author    = {Souza, Ivo and Marzari, Nicola and Vanderbilt, David},
  journal   = {Phys. Rev. B},
  title     = {Maximally localized Wannier functions for entangled energy bands},
  year      = {2001},
  month     = {Dec},
  pages     = {035109},
  volume    = {65},
  doi       = {10.1103/PhysRevB.65.035109},
  issue     = {3},
  numpages  = {13},
  publisher = {American Physical Society},
  url       = {https://link.aps.org/doi/10.1103/PhysRevB.65.035109},
}

@Article{9,
  author    = {Franchini, C. and Kováčik, R. and Marsman, M. and Sathyanarayana Murthy, S. and He, J. and Ederer, C. and Kresse, G.},
  journal   = {J. Phys.: Condens. Matter},
  title     = {Maximally localized Wannier functions in {LaMnO$_{3}$} within {PBE} + {U}, hybrid functionals and partially self-consistent GW: an efficient route to construct ab initio tight-binding parameters for eg perovskites},
  year      = {2012},
  issn      = {0953-8984},
  number    = {23},
  pages     = {235602},
  volume    = {24},
  doi       = {10.1088/0953-8984/24/23/235602},
  publisher = {IOP Publishing},
  url       = {https://dx.doi.org/10.1088/0953-8984/24/23/235602},
}

@Article{3,
  author    = {Perdew, John P. and Burke, Kieron and Ernzerhof, Matthias},
  journal   = {Phys. Rev. Lett.},
  title     = {Generalized Gradient Approximation Made Simple},
  year      = {1996},
  month     = {Oct},
  pages     = {3865--3868},
  volume    = {77},
  doi       = {10.1103/PhysRevLett.77.3865},
  issue     = {18},
  numpages  = {0},
  publisher = {American Physical Society},
  url       = {https://link.aps.org/doi/10.1103/PhysRevLett.77.3865},
}

@Article{4,
  author    = {Kresse, G. and Joubert, D.},
  journal   = {Phys. Rev. B},
  title     = {From ultrasoft pseudopotentials to the projector augmented-wave method},
  year      = {1999},
  month     = {Jan},
  pages     = {1758--1775},
  volume    = {59},
  doi       = {10.1103/PhysRevB.59.1758},
  issue     = {3},
  numpages  = {0},
  publisher = {American Physical Society},
  url       = {https://link.aps.org/doi/10.1103/PhysRevB.59.1758},
}

@Article{66,
  author    = {Baroni, Stefano and de Gironcoli, Stefano and Dal Corso, Andrea and Giannozzi, Paolo},
  journal   = {Rev. Mod. Phys.},
  title     = {Phonons and related crystal properties from density-functional perturbation theory},
  year      = {2001},
  month     = {Jul},
  pages     = {515--562},
  volume    = {73},
  doi       = {10.1103/RevModPhys.73.515},
  issue     = {2},
  numpages  = {0},
  publisher = {American Physical Society},
  url       = {https://link.aps.org/doi/10.1103/RevModPhys.73.515},
}

@Article{67,
  author    = {Giustino, Feliciano},
  journal   = {Rev. Mod. Phys.},
  title     = {Electron-phonon interactions from first principles},
  year      = {2017},
  month     = {Feb},
  pages     = {015003},
  volume    = {89},
  doi       = {10.1103/RevModPhys.89.015003},
  issue     = {1},
  numpages  = {63},
  publisher = {American Physical Society},
  url       = {https://link.aps.org/doi/10.1103/RevModPhys.89.015003},
}

@Article{Dirac,
  author    = {Low, Achintya and Bhowmik, Tushar Kanti and Ghosh, Susanta and Thirupathaiah, Setti},
  journal   = {Phys. Rev. B},
  title     = {Anisotropic nonsaturating magnetoresistance observed in {H}o{M}n$_{6}${G}e$_{6}$: A kagome Dirac semimetal},
  year      = {2024},
  month     = {May},
  pages     = {195104},
  volume    = {109},
  doi       = {10.1103/PhysRevB.109.195104},
  issue     = {19},
  numpages  = {9},
  publisher = {American Physical Society},
  url       = {https://link.aps.org/doi/10.1103/PhysRevB.109.195104},
}

@Article{Flatband1,
  author   = {Neves, Paul M. and Wakefield, Joshua P. and Fang, Shiang and Nguyen, Haimi and Ye, Linda and Checkelsky, Joseph G.},
  journal  = {npj Comput. Mater.},
  title    = {Crystal net catalog of model flat band materials},
  year     = {2024},
  issn     = {2057-3960},
  number   = {1},
  pages    = {39},
  volume   = {10},
  doi      = {10.1038/s41524-024-01220-x},
  refid    = {Neves2024},
  url      = {https://doi.org/10.1038/s41524-024-01220-x},
}

@Article{VHS2,
  author   = {Scammell, Harley D. and Ingham, Julian and Li, Tommy and Sushkov, Oleg P.},
  journal  = {Nat. Commun.},
  title    = {Chiral excitonic order from twofold van Hove singularities in kagome metals},
  year     = {2023},
  issn     = {2041-1723},
  number   = {1},
  pages    = {605},
  volume   = {14},
  doi      = {10.1038/s41467-023-35987-2},
  refid    = {Scammell2023},
  url      = {https://doi.org/10.1038/s41467-023-35987-2},
}

@Article{VHS3,
  author   = {Hu, Yong and Wu, Xianxin and Ortiz, Brenden R. and Ju, Sailong and Han, Xinloong and Ma, Junzhang and Plumb, Nicholas C. and Radovic, Milan and Thomale, Ronny and Wilson, Stephen D. and Schnyder, Andreas P. and Shi, Ming},
  journal  = {Nat. Commun.},
  title    = {Rich nature of Van Hove singularities in Kagome superconductor {C}s{V}$_{3}${S}b$_{5}$},
  year     = {2022},
  issn     = {2041-1723},
  number   = {1},
  pages    = {2220},
  volume   = {13},
  doi      = {10.1038/s41467-022-29828-x},
  refid    = {Hu2022},
  url      = {https://doi.org/10.1038/s41467-022-29828-x},
}

@Article{VHS1,
  author    = {Hu, Yong and Wu, Xianxin and Yang, Yongqi and Gao, Shunye and Plumb, Nicholas C. and Schnyder, Andreas P. and Xie, Weiwei and Ma, Junzhang and Shi, Ming},
  journal   = {Sci. Adv.},
  title     = {Tunable topological Dirac surface states and van Hove singularities in kagome metal {G}d{V}$_{6}${S}n$_{6}$},
  year      = {2025},
  month     = oct,
  number    = {38},
  pages     = {eadd2024},
  volume    = {8},
  comment   = {doi: 10.1126/sciadv.add2024},
  doi       = {10.1126/sciadv.add2024},
  publisher = {American Association for the Advancement of Science},
  url       = {https://doi.org/10.1126/sciadv.add2024},
}

@Article{xiaopengcheng,
  author    = {Xiao, Peng Cheng and Yang, Liu and Lu, Hong Yan and Hao, Ning and Zhang, Ping},
  journal   = {Phys. Rev. B},
  title     = {Prediction of a kagome topological superconducting family: {X}{B}$_{3}$ ({X}={Ni}, {Pd})},
  year      = {2024},
  month     = {Feb},
  pages     = {054506},
  volume    = {109},
  doi       = {10.1103/PhysRevB.109.054506},
  issue     = {5},
  numpages  = {9},
  publisher = {American Physical Society},
  url       = {https://link.aps.org/doi/10.1103/PhysRevB.109.054506},
}

@article{2020PRB-MnB3,
  title = {Boron kagome-layer induced intrinsic superconductivity in a {M}n{B}$_{3}$ monolayer with a high critical temperature},
  author = {Qu, Ziyang and Han, Fanjunjie and Yu, Tong and Xu, Meiling and Li, Yinwei and Yang, Guochun},
  journal = {Phys. Rev. B},
  volume = {102},
  issue = {7},
  pages = {075431},
  numpages = {6},
  year = {2020},
  month = {Aug},
  publisher = {American Physical Society},
  doi = {10.1103/PhysRevB.102.075431},
  url = {https://link.aps.org/doi/10.1103/PhysRevB.102.075431}
}

@article{PRBFeSn-AFM,
  title = {Reorientation of antiferromagnetism in cobalt doped {F}e{S}n},
  author = {Meier, William R. and Yan, Jiaqiang and McGuire, Michael A. and Wang, Xiaoping and Christianson, Andrew D. and Sales, Brian C.},
  journal = {Phys. Rev. B},
  volume = {100},
  issue = {18},
  pages = {184421},
  numpages = {7},
  year = {2019},
  month = {Nov},
  publisher = {American Physical Society},
  doi = {10.1103/PhysRevB.100.184421},
  url = {https://link.aps.org/doi/10.1103/PhysRevB.100.184421}
}

\end{document}